\documentclass{article}

\usepackage{authblk}
\usepackage{latexsym}
\usepackage{amssymb}
\usepackage{url}
\usepackage{colortbl}

\def\false{\mbox{\textit{false}}}
\def\glb{\mbox{\textit{Glb}}}

\def\img{\mbox{\textit{Img}}}

\def\rng{\mbox{\textit{Rng}}}
\def\simp{\mbox{\textnormal{\sf Simp}}}

\def\true{\mbox{\textit{true}}}

\newtheorem{definition}{Definition}[section]

\newtheorem{example}{Example}[section]

\newtheorem{proposition}[definition]{Proposition}

\title{Don't go gaga with GIGO}

\author{\vspace{-1.5cm}}

\date{}

\begin{document}

\maketitle

\begin{center}
\begin{tabular}{c c}
\textbf{Hendrik Decker} & \textbf{Davide Martinenghi} \\
\textit{LMU Munich, Germany} & \textit{Politecnico di Milano, Italy} \\
\texttt{hdecker@pms.ifi.lmu.de} & \texttt{davide.martinenghi@polimi.it}
\end{tabular}
\end{center}

\begin{abstract}
We revisit integrity checking in relational and deductive databases with an approach that tolerates erroneous, inconsistent data.
In particular, we relax the fundamental prerequisite that, in order to apply any method for simplified integrity checking, all data must initially have integrity.
As opposed to a long-standing belief, integrity in the old state before the update is not needed for a correct
application of simplification methods.
Rather, we show that correct simplifications preserve what was consistent across updates.
We formally characterize this property, that we call inconsistency tolerance, and state its validity for some well-known methods for integrity checking.
\end{abstract}

\section{Introduction}\label{sec:intro}

Integrity checking has been a perennial topic in almost all database conferences, journals and
research labs.
The motivation
behind this
is that integrity checking
is practically unfeasible for significant amounts of stored data
without a dedicated approach to optimize the process.
Progress has been made with extensions of basic approaches
in several database areas.
However,
the fundamental ideas that are already present in the seminal paper \cite{Nic82} have not changed much.
The basic principle is that, in most cases, a simplified form of the set
of integrity constraints imposed on the database can be obtained
from a given update (or just an update schema) and the current state
of the database (or just the database schema). Thus, integrity,
which is supposed to be an invariant of all possible database
states, is checked upon each update request, which in turn is authorized only 
if the simplified check yields that integrity is not violated. Here,
``simplified'' essentially means ``more efficiently evaluated at update
time''.

The ``Garbage-In, Garbage-Out'' (GIGO) problem may show up in a database whenever it contains information that is considered erroneous or unwanted, i.e., data that violate the integrity constraints. Then, answers to queries cannot be trusted, since they may be (partly or totally) composed of tuples coming from such erroneous data.
This problem has typically been addressed in simplification methods to integrity checking in a very drastic way: in order to avoid ``Garbage-Out'' (possibly wrong answers), ``Garbage-In'' (possibly incorrect data in the database tables) needs to be completely prevented, even for the applicability of the simplification method itself.
But this is most often unrealistic: the total absence  of unwanted, incorrect or unexpected data is definitely an exception in virtually all real-world scenarios. Still, it is desirable to preserve the ``good'' data in the database while preventing further garbage from sneaking in and, thus, diminish trustability of query answers. 

More formally, in \cite{Nic82} and in virtually all publications on the same subject that came after it, a categorical premise for the correctness of the simplification approach has been that the constraints to be checked for a given update $U$ are satisfied in the ``old" state, i.e., the database state given when $U$ is requested. Otherwise, correctness of simplification is not guaranteed.

As opposed to the attention granted to integrity checking in
academia, support for the declarative specification and efficient
evaluation of semantic integrity in practical systems has always
been relatively scant, apart from standard constructs such as
constraints on column values and foreign keys in relational database
tables. Various reasons have been identified for this 
lack of practical attention. Among them, the logically abstract
presentation of many of the known simplification and 
the possible loss of performance
due to integrity checking are often mentioned.
Here, we focus on another
issue of integrity checking which is responsible for a severe
mismatch between theory and practice:
hardly any database ever is in a perfectly consistent
state with regard to its intended semantics.
Clearly, this contradicts the fundamental premise that
the database must always satisfy integrity. Thus, due to the
intolerance of classical logic wrt inconsistency, integrity checking is very often not considered
an issue of practical feasibility or even relevance.

In this paper, we argue that
inconsistency is far less harmful for database integrity than as
suggested by commonly established results. We substantiate our claim
by showing that, informally speaking, the consistent part of a
possibly inconsistent database can be preserved across updates. More precisely, we
show that, if the simplified form of an integrity theory is
satisfied, then each instance of
each constraint that has been satisfied in the old state continues
to be satisfied in the new state, even if the database is not fully consistent.
Such an approach is therefore tolerant wrt the presence of ``Garbage-In'' and yet helps preventing occurrences of ``Garbage-Out''. 

\section{Integrity checking, simplified}
\label{sec:integrity-constraints-simplified}

Throughout, we refer to the relational framework of deductive
databases,
i.e., relational databases with possibly recursive view definitions
described in clause form \cite{AHV95}.
Thus, a {\em database} consists of a set of \emph{facts}
and a set of \emph{rules}, i.e.,
\emph{tuples} and \emph{view} definitions,
respectively, in the terminology of the relational model.

An {\em integrity constraint} (IC) expresses a semantic invariant, i.e.,
a condition supposed to hold in each state of the database. In
general, it can be expressed by any closed first-order logic formula
in the language of the database on which it is imposed.
Usually, w.l.o.g., ICs are either represented in prenex normal form or as denials, the former with all quantifiers moved leftmost and all negation symbols moved innermost, the latter as datalog clauses with empty head, expressing that, if their condition
is satisfied, then integrity is violated.
An \emph{integrity theory}
is a finite set of ICs.

We limit ourselves to databases with a
unique standard model.
For a closed formula $W$, we write $D\models W$ (resp.,
$D\not\models W$) to indicate that $W$ evaluates to $\true$ (resp.,
$\false$) in $D$'s standard model.
For a set of formulas $\Gamma$, we write $D\models \Gamma$ (resp.,
$D\not\models \Gamma$) to indicate that for every (resp., some)
formula $W\in\Gamma$ we have $D\models W$ (resp., $D\not\models W$).
If $W$ is an IC and $\Gamma$ an integrity theory,
it is usual to also say that $D$ {\em satisfies} (resp., {\em
violates}) $W$ and $\Gamma$, respectively. Also, $D$ is said to be {\em consistent} with an integrity theory $\Gamma$
iff $D\models\Gamma$.

Database states are determined by atomically executed
updates. An \emph{update} $U$ is a mapping $U:{\cal D} \mapsto {\cal
D}$, where ${\cal D}$ is the space of databases
as determined by a
fixed language.
For convenience, for any database $D$,
we denote with $D^U$ the new database
obtained by
applying update $U$ on $D$.

The integrity checking problem asks, given a set of ICs $\Gamma$, a database $D$ consistent with $\Gamma$, and
an update $U$,
whether $D^U\models\Gamma$ holds, i.e., whether the new database satisfies the ICs.
However,
evaluating $\Gamma$ in $D^U$ may be prohibitively expensive.
So, a reformulation of the problem is called for, trying to take
advantage of the incrementality of updates. Typically, all such
reformulations are made
under the assumption that the old state
is consistent.

We will discuss two kinds of
such reformulations
that are common in the literature.
Both determine an alternative integrity theory, the evaluation of which is
supposed to be simpler than to evaluate $\Gamma$, while yielding
equivalent results.
The first kind
of such theory
is determined to be evaluated in the new
state.
\begin{definition}[Post-test]\label{def:post-test}
Let $\Gamma$ be an integrity theory and $U$ an update. An integrity
theory $\Upsilon$ is a \emph{post-test} of $\Gamma$ for $U$ whenever
$D^U\models\Gamma\mbox{ iff } D^U\models \Upsilon$ for
every
database $D$ consistent with $\Gamma$.
\end{definition}
Clearly, $\Gamma$ itself is a post-test of $\Gamma$
for
any update.
But, as indicated, one is interested in producing a post-test that is
actually ``simpler'' to evaluate than the original ICs. This
is traditionally
achieved by exploiting
the fact
that the old
state $D$ is consistent with $\Gamma$, thus
avoiding redundant checks
of cases that are already known to satisfy integrity.
Note that the process of integrity checking involving a post-test
consists in: executing the update, checking the post-test and, if it
fails to hold, correcting the database by
performing a repair action, i.e., a rollback
and optionally a modification of the update
which won't violate integrity.
Approaches of this kind are described in
\cite{Nic82,LST87,LD98}.

The second kind of approach to deal with integrity checking
incrementally is to determine an integrity theory $\Sigma$
to be evaluated in the old state, i.e., to predict without actually
executing the update whether the new, updated state will be
consistent with the ICs.
\begin{definition}[Pre-test]\label{def:pre-test}
Let $\Gamma$ be an integrity theory and $U$ an update. An integrity
theory $\Sigma$ is a \emph{pre-test} of $\Gamma$ for $U$ whenever
$D^U\models\Gamma\mbox{ iff } D\models \Sigma$ for
every
database $D$ consistent with $\Gamma$.
\end{definition}
Here, not only is the consistency of the old state $D$ with $\Gamma$ 
exploited, but also the fact that inconsistent states can be prevented
without executing the update, and, thus, without ever
having to undo a violated new state.
The integrity checking process involving a pre-test is therefore:
check whether the pre-test holds and, if it does, execute the
update. Examples of pre-test-based approaches are
\cite{Qia88,CM:FI2006}.

We refer to both post- and pre-tests
as \emph{simplifications} of the original integrity theory.
In general, 
it is desirable to have kindred
reference pre- and post-tests for benchmarking given
simplifications\footnote{We note that the evaluation costs of pre-tests and post-tests cannot be easily compared with one another
since they refer to two different (viz. old and new) states.
Yet, post-tests can always be compared with the original ICs, since they are meant to be checked on the same state (the updated database).}.
Plain tests, as defined below, i.e., simplifications that do not exploit
the fact that the old state has integrity, and thus are ``non-optimized'', 
may serve as such reference tests.
\begin{definition}[Plain test]\label{def:plain-test}
Let $\Gamma$ be an integrity theory and $U$ an update.

a) An integrity theory $\Sigma_0$ is a \emph{plain pre-test} of
$\Gamma$ for
$U$ if the following holds:\\
$D\models\Sigma_0 \mbox{
iff } D^U\models\Gamma$ for every database $D$.

b) An integrity theory $\Upsilon_0$ is a \emph{plain post-test} of
$\Gamma$ for $U$ if the following
holds:\\
 $D^U\models\Upsilon_0 \mbox{ iff } D^U\models\Gamma$  for
every database $D$.
\end{definition}
Clearly, $\Gamma$ is a plain
post-test
 of itself
for
any update.
Since each 
plain test is a simplification,
it is desirable that ``good'' simplifications be at least as simple
to evaluate
as any corresponding plain test.

We conclude with an example of simplification.
\begin{example}\label{ex:conflict-of-interests}
Consider a database with the relations $rev(S,R)$ (submission $S$
assigned to reviewer $R$), $sub(S,A)$ (submission $S$ authored by
$A$) and $pub(P,A)$ (publication $P$ authored by $A$). Assume a
policy establishing that no one can review a paper by him/herself or a
(possibly former) coauthor. This is expressed by:
$$\begin{array}{rl}
\!\!\Gamma\!\!=\!\!\{&\!\!\!\!\leftarrow \!\!rev(S,R)\land sub(S,R),\\
\!\!& \!\!\!\!\leftarrow \!\!rev(S,R)\land sub(S,A)\land
pub(P,R)\land pub(P,A)\}
\end{array}$$
Let $U$ be the insertion of the facts $sub(c,a)$ and
$rev(c,b)$ into the database, where $a$, $b$, $c$ are some
constants.
A simplification of $\Gamma$ for $U$ is as follows:
$$\begin{array}{rl}
\Sigma=\{\!\!\!\!\!\!\!\!&\leftarrow sub(c,b), \mbox{ [$b$ did not submit $c$]}\\
& \leftarrow rev(c,a), \mbox{ [$a$ does not review $c$]}\\
& \leftarrow pub(P,b)\land pub(P,a), \mbox{ [$b$ is not coauthor of $a$]}\\
& \leftarrow sub(c,A)\land pub(P,b)\land pub(P,A),\\
& \mbox{ \hspace{1cm}[$b$ is not coauthor of an author of $c$]}\\
& \leftarrow rev(c,R)\land pub(P,R)\land pub(P,a)\;\;\}\\
& \mbox{ \hspace{1cm}[$c$ is not reviewed by a coauthor of $a$]}
\end{array}$$
These checks are arguably easier to execute than $\Gamma$ (as well as any other plain test), as they
greatly reduce the space of tuples to be considered by virtue of the
instantiation of variables with constants.
\end{example}

\section{Integrity checking, tolerant}\label{sec:inconsistencyTolerantSimplification}

In order to define inconsistency tolerance for simplifications,
we need to introduce the notion of ``case" of an IC, which formalizes the thought that integrity
typically is violated not as a whole, but only by cases. In this way
it is possible to deal with consistent (satisfied) cases
of ICs in isolation from inconsistent (violated) cases.
To clarify
this, we employ the notion of \emph{substitution}, i.e., a mapping from variables to terms. A substitution $\sigma$ may also be written as
$\{X_{1},\dots,X_{n}/t_{1},\dots,t_{n}\}$ to indicate that each variable $X_{i}$ is
orderly mapped to term $t_{i}$; the notation
$\rng(\sigma)$ refers to the set of variables $X_{i}$, $\img(\sigma)$ to the set of variables among the terms $t_{i}$.
Whenever $E$ is a term or formula and $\sigma$ is a
substitution,
$E\sigma$ is called an
\emph{instance} of $E$.
A variable $x$ occurring in an IC $W$ is a \emph{global variable} in $W$ if it is $\forall$-quantified but
not dominated by any $\exists$ quantifier (i.e., $\exists$ does not
occur left of the quantifier of $x$ in $W$) in the prenex normal
form of $W$;
$\glb(W)$ denotes the set of global variables in $W$.

\begin{definition}[Case]
\label{def:case-of-integrity-constraint}
Let $W$ be an IC.
Then 
$W\sigma$ is called a {\em case of} $W$ if
$\sigma$ is a substitution s. t.
$\rng(\sigma)$\,$\subseteq$\,$\glb(W)$ and
$\img(\sigma)$\,$\cap$\,$\glb(W)$\,\,$=$\,\,$\emptyset$.
\end{definition}
Clearly, each variable in an IC 
 $W$ represented
as a denial
is a global variable of $W$.
Note that cases of an IC need not be ground,
and that each IC $W$ as well as each variant of $W$ is a
case of $W$.

\begin{definition}[Inconsistency tolerance]\label{def:inco-tol}
Let $\Gamma$ be an integrity theory, $U$ an update and $\phi$ a case
of an IC in $\Gamma$. A pre-test $\Sigma$ (resp., post-test
$\Upsilon$) of $\Gamma$ for $U$ is \emph{inconsistency-tolerant}
whenever $D^U\models\phi \mbox{ if } D\models\Sigma$ (resp.,
$D^U\models\phi \mbox{ if } D^U\models\Upsilon$) for all
$D$ s.t. $D\models\phi$.
\end{definition}
In order to characterize simplification methods with respect to
inconsistency tolerance, let us indicate with
$\simp_{\cal M}^U(\Gamma)$ the simplification returned by
method $\cal M$ on input integrity theory $\Gamma$
and input update $U$.
\begin{definition}
A simplification method $\cal M$ is inconsistency-tolerant if
$\simp_{\cal M}^U(\Gamma)$ is an inconsistency-tolerant (pre- or post-) test
of $\Gamma$ for $U$ for any integrity theory $\Gamma$ and
update $U$.
\end{definition}
\begin{example}[\ref{ex:conflict-of-interests} cont'd]
Suppose that the facts $rev(d,e)$ and $sub(d,e)$ are in $D$. Clearly, $D\not\models\Gamma$. However, $U$ is not going to introduce new violations of any IC in $\Gamma$ as long as the test 
$\Sigma$ obtained in example \ref{ex:conflict-of-interests} succeeds.
Informally, the method that returned $\Sigma$ tolerates inconsistency in $D$ and can be used to guarantee that all the cases of $\Gamma$ that were satisfied in $D$ will still be satisfied in $D^{U}$.
\end{example}

We first note that inconsistency tolerance is not satisfied in
general, which immediately limits our ambition of possibly
characterizing all conceivable simplification methods as
inconsistency-tolerant.
\begin{example}\label{ex:not-all-inco-tol}
Consider $D = \{p(a)\}$, $\Gamma = \{\leftarrow p(X)\}$, $\phi
=\leftarrow p(b)$ and let $U$ be the insertion of $p(b)$. 
Integrity theory $\Upsilon=\exists X (p(X)\land X\neq b)$ is a
post-test of $\Gamma$ for $U$, since, whenever $\Gamma$ holds,
$\Upsilon$ evaluates to $\false$, which it should, since $U$
invalidates $\Gamma$. With a similar argument we can conclude that
$\Sigma = \exists X p(X)$ is a pre-test of $\Gamma$ for $U$. However, we have
$D\models\Sigma$ and $D^U\models\Upsilon$, although $D\models\phi$ and $D^U\not\models\phi$, so neither $\Sigma$ nor $\Upsilon$ are inconsistency-tolerant.
\end{example}
Yet, it is easily seen that inconsistency tolerance is enjoyed by plain pre-tests and
post-tests.
\begin{proposition}\label{pro:plain-is-inconsistency-tolerant}
Let $\Gamma$ be an integrity theory, $U$ an update and $\phi$
a case of an IC in $\Gamma$. Any plain pre-test (resp., plain post-test) of $\Gamma$ for $U$ is an inconsistency-tolerant pre-test
(resp., post-test) of $\Gamma$ for $U$.
\end{proposition}
As for concrete integrity checking methods, we have verified that the approaches in \cite{Nic82,LST87} and variations thereof that do not make cross-constraint optimizations are indeed inconsistency-tolerant, whereas the method of \cite{CM:FI2006} is guaranteed to be tolerant only if applied to integrity theories consisting of a single IC.
These results provide us with tools that can have a major practical impact, namely that the ``measure'' of inconsistency, in a sense that can be defined, e.g., as in \cite{Grant:2006fk}, cannot increase as long as an inconsistency-tolerant approach is used.
For example, in a relational database, the percentage of the data that
participate in inconsistencies will necessarily decrease in the new
state if the update consists only of insertions, as is the case for
example \ref{ex:conflict-of-interests} (the simplification of which would be produced by any of the cited inconsistency-tolerant methods); 
in general, preventing the introduction of new inconsistency will improve integrity.
Take, e.g., a federation of databases: initially,
a fair amount of inconsistency can be expected.  For instance, a
business which has taken over a former competitor possibly cannot
wait to continue to operate with a new database resulting from
federating the overtaken one with its own, despite
inconsistencies. However, with an
inconsistency-tolerant integrity checking method, the amount of inconsistency
will decrease over time.

\section{Related work}\label{sec:related}

Simplification of integrity constraints has been recognized by a large body of research as a powerful technique for optimization of integrity checking.
Several approaches to simplification
require the update transaction to be performed
{\em before} the resulting state is
checked for consistency with a
post-test~\cite{Nic82,LST87,GM90,DC94,LD98}.
As opposed to that,
pre-tests allow for avoiding both the execution of the
update and, particularly, the restoration of the database state
before the update, which may be very costly.
Pre-test-based methods are, e.g.,
\cite{HMN84,DBLP:conf/sigmod/HsuI85,Qia88,CM:FI2006,DBLP:conf/lopstr/ChristiansenM03,DBLP:conf/foiks/ChristiansenM04,DBLP:conf/lpar/ChristiansenM05,DBLP:conf/adbis/MartinenghiC05,DBLP:conf/dexaw/ChristiansenM06}.

For an account of the cited methods, we refer to surveys on the subject, e.g., \cite{MCD}. Here, we only mention that integrity checking has been often rephrased in terms of other problems, namely \emph{partial evaluation} \cite{LD98}, \emph{materialized view maintenance}~\cite{310709,DBLP:journals/sigmod/DongS00}, and \emph{query containment} \cite{CM:FI2006}.

Logic programming-based approaches such as \cite{SK88} do not
take into account irrelevant clauses for refuting denial
constraints, even if they would take part in an unnoticed case of
inconsistency that has not been caused by the checked update but by
some earlier event. Moreover, such approaches do not exhibit any
explosive behavior as predicted classical logic in the presence of
inconsistency. In other words, query evaluation procedures based on
SL-resolution can fairly well be called inconsistency-tolerant or
``paraconsistent" in a procedural sense, as done, e.g., in
\cite{kowalski_book}.
The declarative inconsistency
tolerance of simplifications for improving integrity checking, as
investigated in this paper, has been considered in numerous works, including~\cite{DBLP:conf/tdm/DeckerM06,DBLP:conf/sebd/DeckerM06,DBLP:conf/lpar/DeckerM06,DBLP:conf/dexaw/DeckerM06,DBLP:conf/advis/DeckerM06,DBLP:conf/dexaw/DeckerM07,DBLP:conf/ppdp/DeckerM08,DBLP:books/igi/Erickson09/DeckerM09,DBLP:conf/er/DeckerM09,DBLP:conf/socialcom/GalliFMTN12,DBLP:journals/tkde/DeckerM11}.
Under that perspective, most approaches can be reconsidered in terms of this declarative understanding of inconsistency tolerance.

Once illegal updates are detected, it must be decided how to restore a consistent database.
The described pre-testing approach is based on a complete \emph{prevention} of inconsistencies: integrity of the updated state is checked in the state preceding the update and, whenever an illegal update is detected, it is not executed.
With post-tests, as mentioned, integrity checking is done, by definition, in the state following the update. In this case, upon
inconsistencies, integrity needs to be restored via corrective
actions. Most typically, such action is a rollback, that simply
cancels the effects of the unwanted update and usually requires
costly bookkeeping in order to restore the old state.
In other circumstances, the update can be considered as an operation
that is not completely illegal, but that can only be accepted
provided that other parts of the database be modified. In this case
a repairing action is needed to change, add or delete tuples of the
database in order to satisfy the integrity constraints again. The
obtained database is called a \emph{repair}.
Since the seminal
contribution \cite{DBLP:conf/pods/ArenasBC99}, many authors have studied the problem of
providing consistent answers to queries posed to an inconsistent
database. These techniques certainly add to inconsistency tolerance
in databases, but cannot be directly used to detect inconsistencies
for integrity checking purposes (i.e., by posing integrity
constraints as queries).
Along the same lines, active rules have been considered as a means
to restore a consistent database
\cite{Dec02}.

\section{Discussion}
All correctness results in
the literature about efficient ways to check integrity 
fundamentally rely on the requirement that integrity must be
satisfied ``in the old state", i.e., before the update is executed.
The conflict between this requirement of
consistency and the failure of databases in practice to comply with
it pushed us towards investigating a more tolerant attitude wrt. inconsistency.
The notion of tolerance discussed in this paper greatly extends the applicability of existing IC checking methods having this property and yet requires no additional effort.
Since, as mentioned, integrity checking is often cast to problems such as partial evaluation, materialized view maintenance, query containment, and, query answering, we believe that relevant future work should adapt and extend the notion of inconsistency tolerance to these contexts.

While consistent query answering techniques cannot be directly used for integrity checking, distance measures used in these contexts may be regarded as a starting point for further investigations about appropriate measures of inconsistency to be applied for tolerant integrity checking.
Related to this, alternative (but equally appropriate) definitions of inconsistency tolerance may be considered, as opposed to the one defined in this paper, which was based on the notion of case.

In summary, a continuous application of any correct inconsistency-tolerant
integrity checking method will over time improve the quality of stored
data in terms of their compliance with the given ICs, without any additional cost.
In other words, the GIGO effect related to data inconsistency
can be localized and thus tamed and controlled.

\bibliographystyle{abbrv}

\begin{thebibliography}{10}

\bibitem{AHV95}
S.~Abiteboul, R.~Hull, and V.~Vianu.
\newblock {\em Foundations of Databases}.
\newblock Addison-Wesley, 1995.

\bibitem{DBLP:conf/pods/ArenasBC99}
M.~Arenas, L.~E. Bertossi, and J.~Chomicki.
\newblock Consistent query answers in inconsistent databases.
\newblock In {\em PODS '99}, pages 68--79. ACM Press, 1999.

\bibitem{DBLP:conf/lopstr/ChristiansenM03}
H.~Christiansen and D.~Martinenghi.
\newblock Simplification of database integrity constraints revisited: {A}
  transformational approach.
\newblock In M.~Bruynooghe, editor, {\em Logic Based Program Synthesis and
  Transformation, 13th International Symposium {LOPSTR} 2003, Uppsala, Sweden,
  August 25-27, 2003, Revised Selected Papers}, volume 3018 of {\em Lecture
  Notes in Computer Science}, pages 178--197. Springer, 2003.

\bibitem{DBLP:conf/foiks/ChristiansenM04}
H.~Christiansen and D.~Martinenghi.
\newblock Simplification of integrity constraints for data integration.
\newblock In D.~Seipel and J.~M.~T. Torres, editors, {\em Foundations of
  Information and Knowledge Systems, Third International Symposium, FoIKS 2004,
  Wilhelminenberg Castle, Austria, February 17-20, 2004, Proceedings}, volume
  2942 of {\em Lecture Notes in Computer Science}, pages 31--48. Springer,
  2004.

\bibitem{DBLP:conf/lpar/ChristiansenM05}
H.~Christiansen and D.~Martinenghi.
\newblock Incremental integrity checking: Limitations and possibilities.
\newblock In G.~Sutcliffe and A.~Voronkov, editors, {\em Logic for Programming,
  Artificial Intelligence, and Reasoning, 12th International Conference, {LPAR}
  2005, Montego Bay, Jamaica, December 2-6, 2005, Proceedings}, volume 3835 of
  {\em Lecture Notes in Computer Science}, pages 712--727. Springer, 2005.

\bibitem{CM:FI2006}
H.~Christiansen and D.~Martinenghi.
\newblock On simplification of database integrity constraints.
\newblock {\em Fundamenta Informaticae}, 71(4):371--417, 2006.

\bibitem{DBLP:conf/dexaw/ChristiansenM06}
H.~Christiansen and D.~Martinenghi.
\newblock On using simplification and correction tables for integrity
  maintenance in integrated databases.
\newblock In {\em 17th International Workshop on Database and Expert Systems
  Applications {(DEXA} 2006), 4-8 September 2006, Krakow, Poland}, pages
  569--576. {IEEE} Computer Society, 2006.

\bibitem{Dec02}
H.~Decker.
\newblock Translating advanced integrity checking technology to sql.
\newblock In J.~H. Doorn and L.~C. Rivero, editors, {\em Database integrity:
  challenges and solutions}, pages 203--249. Idea Group Publishing, 2002.

\bibitem{DC94}
H.~Decker and M.~Celma.
\newblock A slick procedure for integrity checking in deductive databases.
\newblock In P.~{Van Hentenryck}, editor, {\em Logic Programming: Proc.\ of the
  11th International Conference on Logic Programming}, pages 456--469. MIT
  Press, Cambridge, MA, 1994.

\bibitem{DBLP:conf/dexaw/DeckerM06}
H.~Decker and D.~Martinenghi.
\newblock Avenues to flexible data integrity checking.
\newblock In {\em 17th International Workshop on Database and Expert Systems
  Applications {(DEXA} 2006), 4-8 September 2006, Krakow, Poland}, pages
  425--429. {IEEE} Computer Society, 2006.

\bibitem{DBLP:conf/sebd/DeckerM06}
H.~Decker and D.~Martinenghi.
\newblock Can integrity tolerate inconsistency?
\newblock In V.~D. Antonellis, C.~Diamantini, and P.~Tiberio, editors, {\em
  Proceedings of the Fourteenth Italian Symposium on Advanced Database Systems,
  {SEBD} 2006, Portonovo (Ancona), Italy, 18-21 June 2006}, pages 32--39, 2006.

\bibitem{DBLP:conf/advis/DeckerM06}
H.~Decker and D.~Martinenghi.
\newblock Checking violation tolerance of approaches to database integrity.
\newblock In T.~M. Yakhno and E.~J. Neuhold, editors, {\em Advances in
  Information Systems, 4th International Conference, {ADVIS} 2006, Izmir,
  Turkey, October 18-20, 2006, Proceedings}, volume 4243 of {\em Lecture Notes
  in Computer Science}, pages 139--148. Springer, 2006.

\bibitem{DBLP:conf/tdm/DeckerM06}
H.~Decker and D.~Martinenghi.
\newblock Integrity checking for uncertain data.
\newblock In A.~de~Keijzer and M.~van Keulen, editors, {\em Second Twente Data
  Management Workshop {(TDM} 2006) on Uncertainty in Databases, Enschede, The
  Netherlands, June 6, 2006}, volume {WP06-01} of {\em {CTIT} Workshop
  Proceedings Series}, pages 41--48. Centre for Telematics and Information
  Technology (CTIT), University of Twente, Enschede, The Netherlands, 2006.

\bibitem{DBLP:conf/lpar/DeckerM06}
H.~Decker and D.~Martinenghi.
\newblock A relaxed approach to integrity and inconsistency in databases.
\newblock In M.~Hermann and A.~Voronkov, editors, {\em Logic for Programming,
  Artificial Intelligence, and Reasoning, 13th International Conference, {LPAR}
  2006, Phnom Penh, Cambodia, November 13-17, 2006, Proceedings}, volume 4246
  of {\em Lecture Notes in Computer Science}, pages 287--301. Springer, 2006.

\bibitem{DBLP:conf/dexaw/DeckerM07}
H.~Decker and D.~Martinenghi.
\newblock Getting rid of straitjackets for flexible integrity checking.
\newblock In {\em 18th International Workshop on Database and Expert Systems
  Applications {(DEXA} 2007), 3-7 September 2007, Regensburg, Germany}, pages
  360--364. {IEEE} Computer Society, 2007.

\bibitem{DBLP:conf/ppdp/DeckerM08}
H.~Decker and D.~Martinenghi.
\newblock Classifying integrity checking methods with regard to inconsistency
  tolerance.
\newblock In S.~Antoy and E.~Albert, editors, {\em Proceedings of the 10th
  International {ACM} {SIGPLAN} Conference on Principles and Practice of
  Declarative Programming, July 15-17, 2008, Valencia, Spain}, pages 195--204.
  {ACM}, 2008.

\bibitem{DBLP:books/igi/Erickson09/DeckerM09}
H.~Decker and D.~Martinenghi.
\newblock Database integrity checking.
\newblock In J.~Erickson, editor, {\em Database Technologies: Concepts,
  Methodologies, Tools, and Applications {(4} Volumes)}, pages 212--220. {IGI}
  Global, 2009.

\bibitem{DBLP:conf/er/DeckerM09}
H.~Decker and D.~Martinenghi.
\newblock Modeling, measuring and monitoring the quality of information.
\newblock In C.~A. Heuser and G.~Pernul, editors, {\em Advances in Conceptual
  Modeling - Challenging Perspectives, {ER} 2009 Workshops CoMoL, ETheCoM,
  FP-UML, MOST-ONISW, QoIS, RIGiM, SeCoGIS, Gramado, Brazil, November 9-12,
  2009. Proceedings}, volume 5833 of {\em Lecture Notes in Computer Science},
  pages 212--221. Springer, 2009.

\bibitem{DBLP:journals/tkde/DeckerM11}
H.~Decker and D.~Martinenghi.
\newblock Inconsistency-tolerant integrity checking.
\newblock {\em {IEEE} Trans. Knowl. Data Eng.}, 23(2):218--234, 2011.

\bibitem{DBLP:journals/sigmod/DongS00}
G.~Dong and J.~Su.
\newblock {Incremental Maintenance of Recursive Views Using Relational
  Calculus/SQL.}
\newblock {\em SIGMOD Record}, 29(1):44--51, 2000.

\bibitem{DBLP:conf/socialcom/GalliFMTN12}
L.~Galli, P.~Fraternali, D.~Martinenghi, M.~Tagliasacchi, and J.~Novak.
\newblock A draw-and-guess game to segment images.
\newblock In {\em 2012 International Conference on Privacy, Security, Risk and
  Trust, {PASSAT} 2012, and 2012 International Confernece on Social Computing,
  SocialCom 2012, Amsterdam, Netherlands, September 3-5, 2012}, pages 914--917.
  {IEEE} Computer Society, 2012.

\bibitem{Grant:2006fk}
J.~Grant and A.~Hunter.
\newblock Measuring inconsistency in knowledgebases.
\newblock {\em Journal of Intelligent Information Systems}, in press.

\bibitem{GM90}
J.~Grant and J.~Minker.
\newblock Integrity constraints in knowledge based systems.
\newblock In H.~Adeli, editor, {\em Knowledge Engineering Vol II,
  Applications}, pages 1--25. McGraw-Hill, 1990.

\bibitem{310709}
A.~Gupta and I.~S. Mumick, editors.
\newblock {\em Materialized views: techniques, implementations, and
  applications}.
\newblock MIT Press, 1999.

\bibitem{HMN84}
L.~Henschen, W.~McCune, and S.~Naqvi.
\newblock Compiling constraint-checking programs from first-order formulas.
\newblock In H.~Gallaire, J.~Minker, and J.-M. Nicolas, editors, {\em Advances
  In Database Theory, February 1-5, 1988, Los Angeles, California, USA},
  volume~2, pages 145--169. Plenum Press, New York, 1984.

\bibitem{DBLP:conf/sigmod/HsuI85}
A.~Hsu and T.~Imielinski.
\newblock Integrity checking for multiple updates.
\newblock In S.~B. Navathe, editor, {\em Proceedings of the 1985 ACM SIGMOD
  International Conference on Management of Data, Austin, Texas, May 28-31,
  1985}, pages 152--168. ACM Press, 1985.

\bibitem{kowalski_book}
R.~A. Kowalski.
\newblock {\em Logic for Problem Solving}.
\newblock Elsevier, 1979.

\bibitem{LD98}
M.~Leuschel and D.~de~Schreye.
\newblock Creating specialised integrity checks through partial evaluation of
  meta-interpreters.
\newblock {\em JLP}, 36(2):149--193, 1998.

\bibitem{LST87}
J.~W. Lloyd, L.~Sonenberg, and R.~W. Topor.
\newblock Integrity constraint checking in stratified databases.
\newblock {\em JLP}, 4(4):331--343, 1987.

\bibitem{DBLP:conf/adbis/MartinenghiC05}
D.~Martinenghi and H.~Christiansen.
\newblock Efficient integrity checking for databases with recursive views.
\newblock In J.~Eder, H.~Haav, A.~Kalja, and J.~Penjam, editors, {\em Advances
  in Databases and Information Systems, 9th East European Conference, {ADBIS}
  2005, Tallinn, Estonia, September 12-15, 2005, Proceedings}, volume 3631 of
  {\em Lecture Notes in Computer Science}, pages 109--124. Springer, 2005.

\bibitem{MCD}
D.~Martinenghi, H.~Christiansen, and H.~Decker.
\newblock Integrity checking and maintenance in relational and deductive
  databases, and beyond.
\newblock In Z.~Ma, editor, {\em Intelligent Databases: Technologies and
  Applications}, pages 238--285. Idea Group Publishing, 2006.

\bibitem{Nic82}
J.-M. Nicolas.
\newblock Logic for improving integrity checking in relational data bases.
\newblock {\em Acta Inf.}, 18:227--253, 1982.

\bibitem{Qia88}
X.~Qian.
\newblock An effective method for integrity constraint simplification.
\newblock In {\em ICDE '88}, pages 338--345. IEEE Computer Society, 1988.

\bibitem{SK88}
F.~Sadri and R.~Kowalski.
\newblock A theorem-proving approach to database integrity.
\newblock In J.~Minker, editor, {\em Foundations of Deductive Databases and
  Logic Programming}, pages 313--362. Morgan Kaufmann, 1988.

\end{thebibliography}

\end{document}